\documentclass[10pt]{iopart}
\usepackage{epsfig}
\begin{document}

\title{Medium effects on the flow of strange particles in heavy ion 
collisions}

\author{Che Ming Ko\dag \footnote[3]{To whom correspondence should be 
addressed (ko@comp.tamu.edu)}}

\address{\dag\ Cyclotron Institute and Physics Department, Texas A\&M 
University, College Station, Texas 77843-3366, USA}

\begin{abstract}
Strange particles such as the kaon and lambda are useful in
probing the properties of dense matter formed in heavy ion
collisions. We review in this talk the theoretical understandings of their 
properties in nuclear medium based on both the effective chiral 
Lagrangian and the phenomenological
hadronic model. We further review the effects due to changes in  
their in-medium properties on their collective flow in
heavy ion collisions.
\end{abstract}


%Uncomment for Submitted to journal title message

%Comment out if separate title page not required
%\maketitle

\section{Introduction}

Strange particle production from heavy ion collisions offers the possibility
to study the properties of the hot dense matter formed in these collisions.
Because of its small interaction cross section with a nucleon,
about 10 mb, Randrup and Ko \cite{randrup1} suggested that 
the kaon would be a good messenger of the
dense matter where it was expected to be produced in heavy ion collisions. 
The first experiment was carried out by Schnetzer 
{\it et al.} \cite {schnetzer} at the Bevalac 
at an energy of 2.1 GeV/nucleon. Although the kaon
yield was consistent with the prediction of cascade calculations,
the momentum spectra of kaons could only be explained if the final-state 
scattering of kaons was included \cite{randrup2,ko1}.  Subsequently, Aichelin
and Ko \cite{aichelin1} showed that the kaon yield in heavy ion collisions
at subthreshold energies, i.e., below 1.56 GeV/nucleon, was sensitive 
to the nuclear equation of state used in the transport model.
Comparisons of the experimental data from the KaoS collaboration at
SIS/GSI \cite{kaos} with results from various transport models 
seem to indicate that the nuclear equation of state is soft
\cite{gqli1,maruyama,hartnack}, consistent with that inferred from the
proton flow data \cite{pflow}.

As to the antikaon, Kaplan and Nelson \cite{kaplan} pointed out,
based on the chiral Lagrangian, that its mass should 
decrease in dense matter due to the attractive vector and scalar potentials,
the latter resulting from the explicit chiral symmetry breaking, 
leading to a possible kaon
condensation in neutron stars \cite{star}. This would limit the
maximum mass of neutron stars to about one and a half solar masses and give
rise to suggestions of many small black holes in our galaxy \cite{black}.
The drop of antikaon mass also gives a natural explanation \cite{gqli2} 
of the observed enhancement
of subthreshold antikaon production in heavy ion collisions by the FOPI
collaboration at SIS/GSI \cite{schroter}. 

Using the relativistic transport model \cite{rvuu}, 
Li {\it et al.} \cite{gqli3} showed that the kaon directed flow in 
heavy ion collisions was sensitive to the strength of kaon potential
in nuclear medium. The observed vanishing kaon flow by the
FOPI collaboration \cite{ritman} was found to be consistent with the
prediction based on the repulsive kaon potential that results
from the cancellation of the attractive
scalar and repulsive vector potentials acting on the kaon. 
Analyses of the the experimental data from heavy ion collisions at
AGS/BNL and SPS/CERN are being carried out, and preliminary results
cannot be explained without including the kaon medium effects in
the transport model. 

In this talk, we will review the theoretical understandings of 
the properties of kaon and lambda in nuclear medium, and the effects 
due to changes in their in-medium properties on their collective
flow in heavy ion collisions at SIS/GSI, AGS/BNL, and SPS/CERN. 

\section{Strange particles in nuclear medium}

\subsection{Kaon and antikaon}
 
Since the pioneering work of Kaplan and Nelson \cite{kaplan}, 
there were many studies on the properties of kaons in dense matter using 
the chiral Lagrangian \cite{chiral}. 
To leading order in the expansion of the chiral Lagrangian,
the explicit symmetry breaking term gives rise to the large kaon
mass, while the term involving the vector current contributes to
its isoscalar $s$-wave scattering amplitude from a nucleon. The
latter leads to a repulsive or an attractive optical potential
for $K^+$ and $K^-$ in symmetric nuclear matter. Further
attraction for the kaon is obtained from terms next to leading
order in the chiral expansion, which involves the kaon-nucleon sigma
term and depends again on the explicit symmetry breaking. Ignoring
isospin dependent terms, which do not contribute in symmetric
nuclear matter, the chiral Lagrangian to the next to leading order is
\begin{eqnarray}
{\cal L_{KN}}&=&-\frac{3}{8f^2}\bar N\gamma_\mu N(\bar K
\buildrel\leftrightarrow\over\partial^\mu K)
+\frac{\Sigma_{KN}}{f^2}\bar NN\bar KK\nonumber\\
&+&\frac{\tilde D}{f^2}(\bar NN)(\partial_\mu\bar K\partial^\mu K).
\end{eqnarray}
In the above, $f\sim 93$ MeV is the pion decay constant, and the
value of the kaon-nucleon sigma-term $\Sigma_{KN} = \frac12 (m_q +
m_s) \langle N | \bar{u} u + \bar{s}s | N \rangle$ depends on the
strangeness content of the nucleon, $y=2\langle N|\bar ss|N\rangle
/\langle N|\bar uu+\bar dd|N\rangle$ $\sim 0.1-0.2$. Using the
light quark mass ratio $m_s/m_{u,d}\sim 29$, one obtains
$370<\Sigma_{KN}<405 \, \rm MeV$. On the other hand, lattice 
gauge calculations show that $y\sim 0.33$ which would give
$\Sigma_{KN}\sim 450$ MeV \cite{lattice}. The additional parameter,
$\tilde{D}\approx 0.33/m_K-\Sigma_{KN}/m_K^2$, is determined from
comparing with $K^+$-nucleon scattering data.

\begin{figure}[htp]
\vspace{0.2in}
\centerline{\epsfig{file=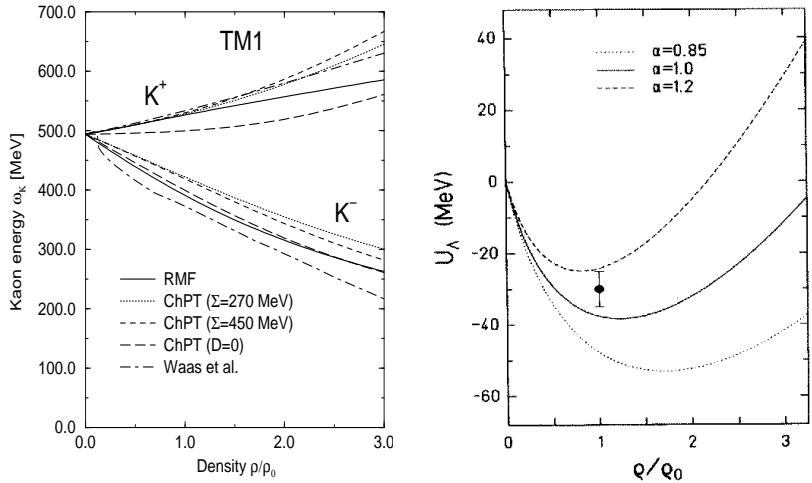,height=2.5in,width=4.2in}}
\caption{Left panel: Kaon and antikaon potential. 
Right panel: Lambda potential for different
values of repulsive vector potential. The empirical value from the
hypernuclear properties is shown by the filled circle.}
\label{klpot}
\end{figure}

In the mean-field approximation, the kaon energy in the nuclear medium 
is then given by
\begin{eqnarray}
\omega_{K,\bar K}&=&\left[\sqrt{(m_K^{*2}+k^2)\left(1+\frac{\tilde D}{f^2}
\rho_s\right)^2+\left(\frac{3}{8f^2}\rho_N\right)^2}
\pm\frac{3}{8f^2}\rho_N\right]\nonumber\\
&\times&\left(1+\frac{\tilde D}{f^2}\rho_s\right)^{-1},
\end{eqnarray}
with
\begin{equation}
m_K^*=\sqrt{\left(m_K^2-\frac{\Sigma_{KN}}{f^2}\rho_s\right)
\left(1+\frac{\tilde D}{f^2}\rho_s\right)^{-1}}.
\end{equation}
In the above, the plus and minus signs are for the kaon and antikaon,
respectively. Using the KFSR relation ($m_\rho=2\sqrt{2}fg_\rho$) and 
the SU(3) relation ($g_\omega=3g_\rho$), the kaon vector
potential, $3\rho_N/(8f^2)$, can be written as
$(1/3)(g_\omega/m_\omega)^2\rho_N$ which is just 1/3 of the nucleon
vector potential in the Walecka model \cite{walecka}, as expected from the
constituent quark model. With $\Sigma_{KN}\sim 350$ MeV, the
in-medium kaon mass at normal nuclear matter density, given by its
energy at zero three-momentum, increases by about 30 MeV,
consistent with that given by the impulse approximation using the
kaon-nucleon scattering length, while that of antikaon decreases by
about 100 MeV, similar to that extracted from the kaonic atom data
\cite{katom}. The kaon and antikaon potential from the chiral Lagrangian 
are shown in the left panel of Fig. \ref{klpot} by the dotted and dashed 
curves for $\Sigma_{KN}$ equal to 270 and 450 MeV, respectively.
 
The chiral Lagrangian predicts, however, an attractive isoscalar 
s-wave scattering length for the $K^-$-nucleon scattering
amplitude, which contradicts with the repulsive one found in experiments 
\cite{martin}. This discrepancy has been attributed to the existence of the 
$\Lambda(1405)$ which is located below the $K^- N$-threshold. 
To circumvent this problem, the chiral perturbation 
calculation was extended either by Lee {\it et al.} \cite{lee}
to include an explicit $\Lambda(1405)$ state or by Kaiser {\it et al.}
\cite{kaiser} to use the interaction obtained from the leading order 
chiral Lagrangian as a kernel for a Lippman-Schwinger type calculation, 
which was then solved to generate a bound state $\Lambda(1405)$.
In this picture, the properties of the $\Lambda(1405)$ is
significantly changed in the nuclear medium
as a result of the Pauli blocking of the proton inside this bound state
\cite{koch}. With increasing density, its mass increases and the
strength of the resonance is reduced, leading to a change of 
the $K^-$ optical potential from repulsive to 
attractive at a density of about $1/4$ of nuclear matter density 
in agreement with the analysis of $K^-$
atoms \cite{katom}. The resulting antikaon potential \cite{waas}
is shown in the left panel of Fig. \ref{klpot} by the dash-dotted curve.

The kaon potential was also studied in the meson-exchange model
by Schaffner {\it et al.} \cite{schaffner1} based on 
the Lagrangian 
\begin{equation}
{\cal L}=\partial_\mu\bar K\partial^\mu K-(m_K^2-g_{\sigma K}
m_K\sigma)\bar KK+ig_{\omega K}\omega_\mu\bar K\buildrel\leftrightarrow 
\over\partial^\mu K,
\end{equation}
where the coupling constants $g_{\sigma K}$ and $g_{\omega K}$ 
to the scalar ($\sigma$) and vector ($\omega$) fields, respectively,
can be related to other known coupling constants using SU(3) symmetry
or determined empirically from the kaon-nucleon scattering amplitude.
In the mean-field approximation, the kaon energy is then given by 
\begin{equation}
\omega_{K,\bar K}=[m_K^2+k^2-g_{\sigma K}m_K\sigma+
(g_{\omega K}\omega_0)^2]^{1/2}\pm g_{\omega K}\omega_0,
\end{equation}
where the plus and minus signs are, again, for the kaon and antikaon,
respectively. The kaon potential given by this model 
is shown in the left panel of Fig. \ref{klpot} by the solid curve.
Although all theoretical models predict that the kaon has a moderate repulsive
potential and the antikaon has a strong attractive potential, their 
magnitudes differ substantially. It is thus important to have experimental
data to constrain the values of the kaon and antikaon potentials. Kaon 
production from heavy ion collisions offers this possibility, particularly
for the kaon potential at high densities.

For the momentum dependence of kaon potential, Shuryak \cite{shuryak} 
found, based on the impulse approximation, 
that it would become weaker as its momentum increases.
On the other hand, the dispersion relation analysis by Sibirtsev and 
Cassing \cite{sibirtsev} shows that the kaon potential is essentially
independent of the momentum. They also found that the 
antikaon potential became less attractive with increasing momentum.
In a coupled channel approach involving $\bar KN$, $\pi\Sigma$, and
$\Lambda(1405)$, Schaffner-Bielich {\it et al.} \cite{schaffner2} 
found that the antikaon potential could even change to a
repulsive one at large momenta due to the diminishing effect
of the Pauli blocking. The latter effect was shown to be 
further enhanced in nuclear matter at finite temperature.

\subsection{Lambda}

Because of strangeness conservation, hyperons such as the lambda and 
sigma are produced together with kaons. 
Heavy ion collisions thus also
offer the opportunity to study the properties of hyperons in dense
matter, which is important in understanding if the core of
a neutron star can exist as a hyperon matter \cite{glen}. 
The potential for the lambda particle at normal nuclear matter density
is relatively well determined
to be about -30 MeV from the structure of hypernuclei
\cite{hyper}. Theoretically, 
the lambda potential in nuclear matter has been studied using
the Dirac-Bruckner-Hartree-Fock approach \cite{dbhf}. 
The result ranges from -25 to -40
MeV, depending on the input boson-exchange models for the $\Lambda N$
interaction. There were also various attempts to generalize the
Walecka-type model from SU(2) to SU(3) to include the hyperon degrees
of freedom \cite{su3}. In the naive SU(3) 
quark model, the hyperon potential is about 2/3 of the nucleon 
potential, as there are only two light quarks in a hyperon
instead of three light quarks in a nucleon.  Recently, hyperon properties 
in nuclear matter were also studied using the QCD sum-rule approach. 
It was found that both lambda scalar and vector potentials were 
significantly weaker than the prediction from the naive quark model 
\cite{qcd1}, while those of the sigma hyperon were somewhat stronger
and were close to the ones for the nucleon \cite{qcd2}. The accuracy of
these findings were, however, limited by uncertainties in the nucleon
strangeness content and certain in-medium four-quark condensates.

Writing the lambda self-energy in terms of the scalar 
($\Sigma_S^\Lambda$) and vector ($\Sigma_V^\Lambda$) parts, 
its potential can then be expressed as
\begin{equation}
U_\Lambda({\bf p},\rho)=[(m_\Lambda-\Sigma_S^\Lambda)^2+{\bf p}^2]^{1/2}
+\Sigma_V^\Lambda-(m_\Lambda^2+{\bf p}^2)^{1/2}.
\end{equation}
Using $\Sigma_S^\Lambda\sim 2\Sigma_S^N/3$ and
$\Sigma_V^\Lambda\sim 2\alpha\Sigma_V^N/3$, where
$\Sigma_S^N$ and $\Sigma_V^N$ are the nucleon scalar and vector 
self-energies, respectively, the lambda potential is shown in 
the right panel of Fig. \ref{klpot}
for different values of $\alpha$. Also shown in the figure is the
empirical lambda potential at normal nuclear matter density.

\section{Strange particle flow in heavy ion collisions}

Strange particle flow in heavy ion collisions
is sensitive to the mean-field potential
in dense matter as an attractive interaction between
the strange particle and nucleons aligns its flow with that of the nucleons 
whereas a repulsion leads to an anti-alignment (anti-flow). However, 
strange particle flow is
also sensitive to the overall reaction dynamics, in particular 
to the properties of the nuclear mean field and to reabsorption processes
especially in case of the antikaons. Therefore, transport calculations 
are required in order to consistently incorporate all these effects. 

\subsection{GSI}

\begin{figure}[htp]
\centerline{\epsfig{file=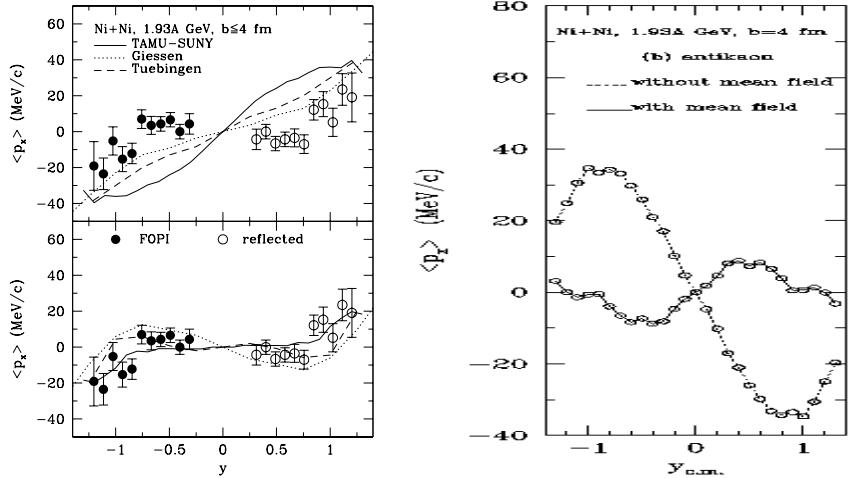,height=2.5in,width=4.4in}}
\caption{Kaon (left panel) and antikaon (right panel) flow.} 
\label{kakflow}
\end{figure} 
 
The $K^+$ flow has been measured by the FOPI collaboration \cite{ritman}
at GSI. Although the error bars are large, the data (circles)
in the left panel of Fig. \ref{kakflow}
clearly show a vanishing flow at midrapidity. Also shown in the figure
are theoretical results from the transport model calculations by
the TAMU-SUNY group \cite{gqli3,gqli4}, the Giessen group \cite{giessen},
and the T\"ubingen group \cite{tubingen}.
It is seen that the experimental data are consistent with the results 
that include the kaon mean-field potential from the chiral Lagrangian 
(lower left panel) but not the ones without any potential (upper left panel). 
It is worthwhile to point out that
while the primordial flow of kaons produced from meson-baryon scattering
is positive, those produced from baryon-baryon collisions already 
have a very small flow \cite{david}. Also, it has been found 
that inclusion of the spatial component of the kaon vector potential 
would lead to an attractive Lorentz force between the kaon and  
nucleons, leading thus again to a positive kaon flow \cite{fuchs}.
However, the Lorentz force increases with kaon momentum, and this
contradicts the predictions from more detailed theoretical studies
which show an opposite momentum dependence \cite{shuryak,sibirtsev}.

\begin{figure}[htp]
\centerline{\epsfig{file=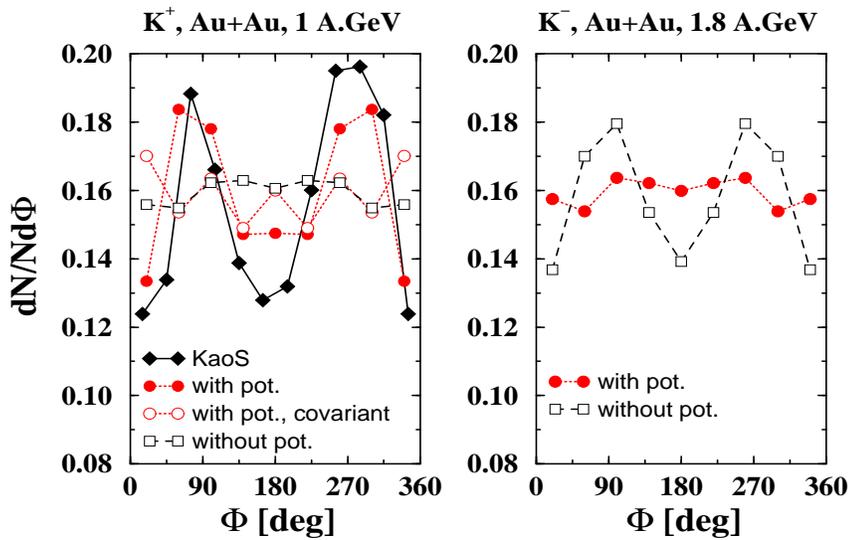,height=2.8in,width=4.4in,angle=0}}
\caption{Azimuthal distributions of $K^+$ (left panel) 
and $K^-$ (right panel) from Ref. \cite{wang1}.}
\label{kphi}
\end{figure}

The repulsive kaon potential also leads to an enhanced emission of
kaons out of the reaction plane as first shown by Li and Ko \cite{gqli5}.
Similar results, shown in the left panel of Fig. \ref{kphi},
were obtained by the T\"ubingen group \cite{wang1}. The observed 
kaon azimuthal distribution, which shows an enhanced out-of-plane
emission, can only be explained by the inclusion of a repulsive
kaon potential. Again, the data do not support the existence of a
simple Lorentz force between the kaon and nucleons.

For antikaons, Li and Ko \cite{gqli6} showed that their 
flow and azimuthal distribution were influenced by the 
attractive potential. 
In the right panel of Fig. \ref{kphi}, the results from the T\"ubingen group 
\cite{wang1} show that the $K^-$ azimuthal distribution,
which is dominated by out-of-plane emission  
in the absence of potential, becomes more symmetric once the
attractive potential is included in the relativistic transport model.
Similarly, the antiflow of antikaons in the absence of potential,
which results from the strong absorption effect, changes into 
a positive flow under the influence of the attractive potential.
The results from Li and Ko are shown in the right panel of Fig. \ref{kakflow}.
Unfortunately, the measurement of $K^-$ observables is more 
difficult since it is produced less abundantly at subthreshold energies.

\begin{figure}[htp]
\vspace{0.2in}
\centerline{\epsfig{file=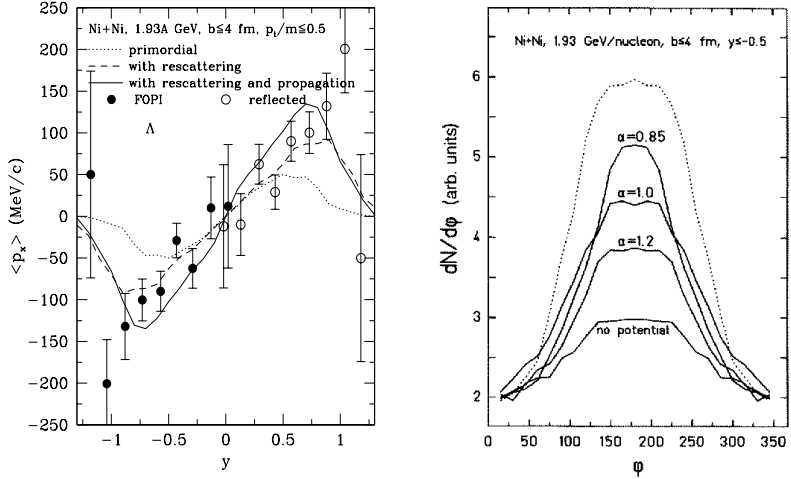,height=2.5in,width=4.1in,angle=0}}
\caption{Lambda flow (left panel) and azimuthal distribution (right panel).}
\label{lambda}
\end{figure}

For lambda particles, which cannot be absorbed as the pion and antikaon, 
it is expected to have a positive flow as the nucleons. Li
and Ko \cite{gqli7} found that the magnitude
of the lambda flow depended on the strength of the attractive
lambda potential. In the left panel of Fig. \ref{lambda}, 
theoretical results from Li and Brown \cite{gqli4}
are shown together with the experimental data from the FOPI collaboration
\cite{ritman}. 
Both the lambda-nucleon scattering and the attractive potential 
are seen to increase the flow. However, the large experimental errors in
the data makes it not possible to draw conclusions on the magnitude
of the lambda potential. A similar conclusion was reached
by the T\"ubingen group \cite{wang2}. Also shown in the right panel of
Fig. \ref{lambda} is the lambda azimuthal distribution predicted
by Li and Ko \cite{gqli7}. It is seen that the lambda potential
has a large effect on the lambda
azimuthal distribution. It will be of interest to 
compare the theoretical predictions with future experimental data.

\subsection{AGS}

\begin{figure}[htp]
\centerline{\epsfig{file=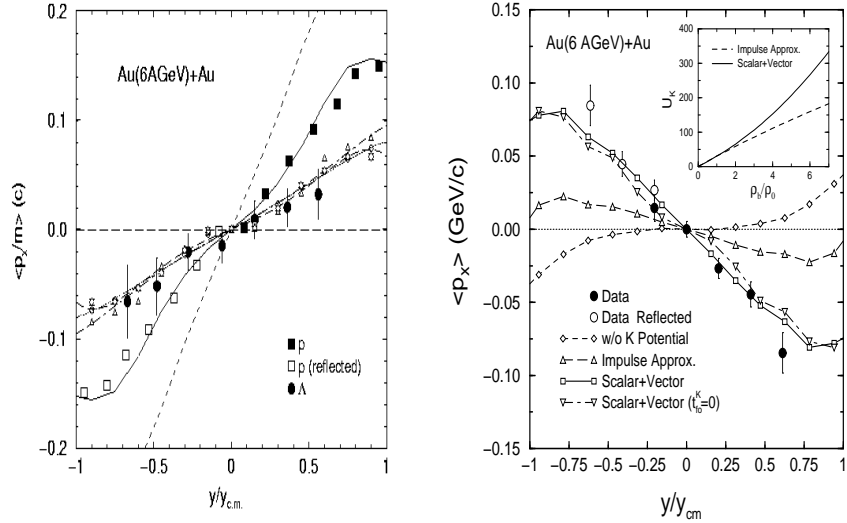,height=3in,width=4.5in,angle=0}}
\vspace{-0.2in}
\caption{Proton and lambda (left panel) as well as kaon (right
panel) directed flow.} 
\label{ags}
\end{figure}

Both lambda and kaon flow were measured in Au+Au collisions
at 6 A GeV at AGS by the E895 Collaboration \cite{e895l,e895k}. 
The data show a positive lambda flow and a large kaon antiflow.
To understand these results, Zhang {\it et al.} \cite{zhang}
and Pal {\it et al.} \cite{pal} have carried out
a calculation based on the ART transport model \cite{art}. 
In the left panel of Fig. \ref{ags}, 
the theoretical results from Ref. \cite{zhang}
are compared with the measured lambda (filled circles)
and proton (squares) flow. The proton flow 
is well described by a soft nuclear equation 
of state corresponding to an incompressibility of 200 MeV (solid curve).
The results from a stiff equation of state with $K=380$ MeV is seen to 
give too large a proton flow. This is different from the conclusion 
based on the UrQMD model \cite{soff}, which includes the string dynamics
in the initial stage, that the stiff equation of state describes the
data better. Although the 
lambda flow obtained with a potential gives a good description of the
data, the large experimental errors cannot differentiate 
the difference between a lambda potential which is 2/3 of the nucleon 
potential (dotted curve) and that which is the same as the nucleon one
(dot-dashed curve). Also shown is the
result obtained with a lambda potential which is 2/3 of the nucleon
potential but without the lambda-nucleon scattering
(dashed curve). Comparing it 
with the dotted curve shows that the scattering effect is not large.

For the kaon flow, the experimental data (circles) show a very large antiflow
compared to that in heavy ion collisions at lower energies available
from GSI.  This is shown in the right panel of Fig. \ref{ags} together with
the theoretical results obtained from the ART model using various
kaon potentials \cite{pal}. Without potential (dashed curve), the kaon has a 
positive flow, which is similar to the results obtained from the RQMD model 
\cite{e895k}. With the potential obtained from the impulse approximation,
shown by the dashed curve in the inset, the kaon flow (dashed curve) 
changes to a small negative one. The results (solid curve) obtained from 
the kaon 
potential given by the chiral Lagrangian, shown by the solid curve in the 
inset, agree very well with the large kaon antiflow measured in experiments.
The latter potential is seen to be more repulsive at high densities
than the one given by the impulse approximation. This is due to
the saturation of the scalar density with increasing densities, which 
makes the attractive scalar kaon potential less important at high
densities than the repulsive vector kaon potential, which increases with 
density. Setting the kaon formation time to zero (${\rm t_{fo}^K=0}$)
(dash-dotted curve) does not affect much the result.
Li {\it et al.} \cite{bali} have pointed out that the kaon flow 
is better illustrated by its dependence on the transverse momentum, 
i.e., the differential flow.  The decrease of kaon flow is
then due to the cancellation between 
the negative flow of kaons with low transverse momenta and the positive 
flow of kaons with high momenta.  

The $K^-$ flow was also studied in the ART model by Song
{\it et al.} \cite{song}
for heavy ion collisions at AGS energies. Similar to the results
at GSI energies, the $K^-$ was predicted
to have a positive flow due to the strong attractive potential.
However, there are at present no experimental data from AGS on
the antikaon flow. 

\subsection{SPS}

For heavy ion collisions at SPS energies, there is very little study
on the flow of strange particles. Experiments by the WA98 collaboration
\cite{wa98} show that the kaon elliptic flow,
which measures the asymmetry between the in-plane and out-of-plane
kaon momentum distributions,
is negative, opposite to that for the proton and the pion. 
This implies that kaons are preferentially emitted out of the reaction 
plane. Theoretical results from 
the RQMD without the kaon potential show instead a positive kaon elliptic 
flow.  Whether the observed negative kaon elliptic flow is due to the
repulsive kaon potential remains to be studied.

\section{Summary}

We have reviewed in this talk the kaon in-medium properties based on the
chiral Lagrangian, which predicts that the kaon feels a moderate repulsive
potential due to the cancellation of the attractive scalar and repulsive
vector potentials while the antikaon has a strong attractive potential
as the vector repulsion becomes attractive for the antikaon. These 
results are consistent with the empirical information obtained from 
the $K^+$-nucleus scattering and the kaonic atom. Similar results are
obtained from the phenomenological meson-exchange model. The latter
model also allows one to study the lambda potential, which is predicted
to have a potential which is somewhat less attractive than that for a nucleon. 

We have also reviewed the experimental data for both kaon
and lambda flow in heavy ion collisions at various energies. For the GSI
energies, both the vanishing kaon flow and enhanced out-of-plane 
emission are consistent with the repulsive
potential predicted by the chiral Lagrangian. The
predicted positive flow of antikaons and their more symmetric azimuthal  
distribution remain to be verified by experiments. For the lambda flow,
which is positive in the experiments, 
the data are not accurate enough to allow for a determination
of the strength of the lambda potential. However, the effect of potential
is predicted to be large in the lambda azimuthal distribution.

For heavy ion collisions at AGS, the large antiflow of kaons observed in the
experiments is again consistent with the kaon potential predicted by
the chiral Lagrangian, which shows that it becomes more repulsive
at high densities than that given by the impulse approximation based
on the kaon-nucleon scattering length. Again, the lambda potential
cannot be determined from the lambda flow data due to the large 
errors in the experiment. Also, there is no data to compare with the 
theoretically predicted positive flow of antikaons.

For heavy ion collisions at SPS, there are
very little experimental measurements
on the flow of strange particles. Preliminary results of a negative
kaon elliptic flow, i,e. out-of-plane emission dominates over
in-plane flow, cannot be accounted for by transport model studies
without a kaon mean-field potential. Further theoretical and
experimental studies are thus needed in order to understand the
properties of strange particles in the hot dense matter formed 
in such high energy heavy ion collisions. 

\section{Acknowledgments}

This work is supported by the National Science Foundation under Grant No. 
PHY-9870038, the Welch Foundation under Grant No. A-1358, and the Texas 
Advanced Research Program under Grant No. FY99-010366-0081.

\end{document}